\documentclass{webofc}
\usepackage[varg]{txfonts}   
\usepackage{graphicx}

\newcommand{\beq}{\begin{equation}}
\newcommand{\eeq}{\end{equation}}
\newcommand{\beqa}{\begin{eqnarray}}
\newcommand{\eeqa}{\end{eqnarray}}

\begin{document}
\title{Constraints on alternative theories of gravity with observations of the Galactic Center}
%
%

\author{\firstname{Alexander} \lastname{Zakharov}\inst{1,2,3,4}\fnsep\thanks{\email{zakharov@itep.ru}} 
}

\institute{Institute of Theoretical and Experimental Physics, 
117218, Moscow, Russia
\and
           Joint Institute for Nuclear Research, Dubna, 141980 Russia
\and
           National Research Nuclear University MEPhI (Moscow Engineering Physics Institute), 
           Moscow, 115409, Russia
\and
North Carolina Central
University, Durham, NC 27707,
 USA
          }

\abstract{%

To evaluate a potential usually one analyzes trajectories of test particles. For the Galactic Center case astronomers use bright stars or photons, so
   there are two basic observational techniques to investigate a gravitational potential, namely, (a) monitoring the orbits of bright stars near the Galactic Center as it is going on with 10m Keck twin  and four 8m VLT telescopes equipped with adaptive optics facilities (in addition, recently
the IR interferometer GRAVITY started to operate with VLT); (b) measuring the size and shape of shadows around black hole with VLBI-technique using telescopes operating in mm-band. At the moment, one can use a small relativistic correction approach for stellar orbit analysis, however, in the future the approximation will not be precise enough due to enormous progress of observational facilities and recently the GRAVITY team found that the first post-Newtonian correction has to be taken into account for the gravitational redshift in the S2 star orbit case. Meanwhile for smallest structure analysis in VLBI observations one really needs a strong gravitational field approximation. We discuss results of observations and their  interpretations.
  In spite of great efforts there is a very slow progress to resolve dark matter (DM) and dark energy (DE) puzzles and in these circumstances in last years a number of alternative theories of gravity have been proposed. Parameters of these theories could be effectively constrained with of observations of the Galactic Center. We show some cases of alternative theories of gravity where their parameters are constrained
with observations, in particular, we consider massive theory of gravity. We choose the alternative theory of gravity since there is a significant activity in this field and in the last years theorists demonstrated an opportunity to create such theories without ghosts, on the other hand, recently, the joint LIGO \& Virgo team presented an upper limit on graviton mass such as $m_g < 1.2 \times 10^{-22}$ eV \cite{Abbott_16} analyzing gravitational wave signal in their first paper where they reported about the discovery of gravitational waves from binary black holes  as it was suggested by C. Will \cite{Will_98}. So, the authors concluded that their observational data do not indicate a significant deviation from classical general relativity.  We show that an analysis of bright star trajectories could estimate a graviton mass with a commensurable accuracy in comparison with an approach  used in gravitational wave observations and the estimates obtained with these two approaches are consistent. Therefore, such an analysis gives an opportunity to treat observations of bright stars near the Galactic Center as a useful tool to obtain constraints on the fundamental gravity law. We showed that in the future graviton mass estimates obtained with analysis of trajectories of bright stars would be better than current LIGO bounds on the value, therefore, based on a potential reconstruction at the Galactic Center we obtain bounds on a graviton mass and these bounds are comparable with LIGO constraints. Analyzing size of shadows around the supermassive black hole at the Galactic Center (or/and in the center of M87) one could constrain parameters of different alternative theories of gravity as well.
}
\maketitle

\section{Observations of bright stars near the Galactic Center}
\label{sec-1}

The closest supermassive black hole is located in the Galactic Center, therefore, this object is very attractive and astronomers observe it in different spectral band including $\gamma$, X-ray, IR, optical and radio.
Moreover, the black hole is a natural laboratory to test general relativity in a strong gravitational field limit.
There are two leading groups observing bright IR stars near the Galactic Center with largest telescopes equipped with adaptive optics facilities.
One group led by A. Ghez uses the twin Keck telescopes with 10 m diameters, another ESO-MPE group led by R. Genzel uses four VLT telescopes with 8 m diameters. Results of these two groups are consistent and complimentary.  Observations showed that stars are moving along elliptical orbits
and therefore, one could conclude that motions of these stars are fitting rather well with a potential of point like mass around $M=4\times 10^6~M_\odot$ in the framework of Newtonian gravity law. One of the most interesting tracer of a gravitational potential at the Galactic Center is S2 star. It has eccentricity $e = 0.88$,  period $T=16$~yr and an expected visible relativistic precession of its orbit is around $\Delta s \approx 0.83$~mas \cite{Gillessen_17} in assumption that extended mass distributions such as a stellar cluster or dark matter near the Galactic Center do not have a significant impact on relativistic precession of its orbit. Currently the Keck uncertainty in the S2 star orbit reconstruction is around $\sigma_{Keck}\approx 0.16$~mas \cite{Hees_PRL_17}, while for Thirty Meter Telescope(TMT) which will be constructed with a several years   $\sigma_{TMT}\approx 0.015$~mas.

\section{GRAVITY in action}

There is a rapid improvement of accuracy of S2 star orbit reconstruction also for MPE--ESO team, since in 1990s a precision of SHARP facilities
were around 4~mas, in 2000s NACO had a precision around 0.5~mas, but the GRAVITY team reached a precision around $30~\mu as$ in 2018 \cite{Gravity_18}.
Analyzing these new GRAVITY data it was shown that GR approach in  post-Newtonian  (PN) approximation provide much better fit in comparison with the Newtonian approach. After observations of S2 star pericenter passage in May 2018 and subsequent data analysis the GRAVITY collaboration reported about the discovery
of general relativity effects for S2 star \cite{Genzel_18,Gravity_18}.
The GRAVITY collaboration evaluated gravitational redshift in the orbit of S2 star near its pericenter passage and relativistic precession of its orbit and showed that observational data are much better fitted with GR model in PN approach than with Newtonian one. It means that almost after 100 years after the confirmation by Dyson,  Eddington and Davidson  of the GR prediction about a deflection of light during Solar eclipse in 1919 \cite{Dyson_20}, astronomers checked GR prediction in much stronger gravitational field at high distances from our Solar system and Einstein's theory of relativity successfully passed one important test more.
Theoretical analysis of gravitational redshift for sources moving in binary system was presented in \cite{Kopeikin_99,Alexander_05,Zucker_06}.
 S2 star passed its pericenter in May 2018 and now it is clear that relativistic corrections have to be taken into account at the period near this passage to fit properly observational data. At the pericenter S2 moves with a total space velocity $V_{\rm peri} \approx 7650$ km/s or $\beta_{\rm peri} = V_{\rm peri}/c=2.55 \times 10^{-2}$ \cite{Gravity_18}.
A total redshift considering the PPN(1) correction could expressed in the following form \cite{Kopeikin_99,Alexander_05,Zucker_06,Gravity_18}
\begin{equation}
 z_{\rm GR}=\frac{\Delta \lambda}{\lambda}=B_0+B_{0.5}\beta+B_1\beta^2+\mathcal{O}(\beta^3),
\label{redshift_1}
\end{equation}
where $B_1=B_{1,tD}+B_{1,grav}$, $B_{1,tD}=B_{1,grav}=0.5$, $B_{1,tD}$ is the special relativistic transverse Doppler effect,
$B_{1,grav}$ is the general relativistic gravitational redshift , $B_{0.5}=\cos \theta$, where $\theta$ is the angle between the velocity vector and line of sight \cite{Alexander_05},  while the total redshift $B_0$ which is independent on a star velocity $\beta$
\begin{equation}
  B_0 = z_\odot + z_{\rm gal}+z_{\rm star} +\frac{1}{2} \Upsilon_0,
  \label{redshift_2}
\end{equation}
therefore the redshift $B_0$ consists of four parts, $z_\odot$ is due a total motion of the Sun and the Earth in respect to Galactic Center and blue shift due to potential of the Sun and the Earth, $z_{\rm gal}$ is redshift due to Galaxy potential, $z_{\rm star}$ is redshift due to the star's potential, the redshift $\dfrac{1}{2}\Upsilon_0 =\dfrac{r_g}{2 r_p}=\dfrac{GM}{c^2 r_p}$ due to the location of star in the SMBH potential \cite{Alexander_05}.
The GRAVITY collaboration evaluated the total redshift from spectroscopical observations and concluded that it corresponds to $z \approx \dfrac{200~{km/s}}{c}$ \cite{Gravity_18}.
One could represent the total redshift obtained from spectroscopical observations in the form \cite{Gravity_18}
\begin{equation}
 z_{\rm tot}=z_K+f(z_{GR}-z_K),
\label{redshift_3}
\end{equation}
where $z_K=B_0+B_{0.5}\beta$ is the Keplerian redshift, $f=0$ corresponds to Keplerian (Newtonian) fit, while $f=1$ corresponds to GR fit describing
with Eq. (\ref{redshift_1}).
The GRAVITY collaboration found that $f=0.90\pm 0.09|_{\rm stat}\pm 0.15 |_{\rm sys}$ and the authors also claimed that S2 data are inconsistent
with the pure Newtonian dynamics. Since $f$-value is slightly less than its expected value estimated with pure GR fit, perhaps an extended mass distribution of stellar cluster should be taken into account in this model and future observations of relativistic redshifts
(and astrometric monitoring the bright stars)
 will
help to evaluate parameters of an extended mass distribution.
The GRAVITY collaboration evaluated also  $f$-value from observational data comparing precessions for Schwarzschild and Newtonian approaches and they concluded that the $f$-value must be much closer to GR value or more precisely $f=0.94 \pm 0.09$ \cite{Gravity_18}.

\section{Evaluations of black hole parameters and constraints on alternative theories of gravity with observations of bright stars near the Galactic Center}

\subsection{Orbital precession due to general central-force
perturbations}

If we assume that a spherically symmetrical Newtonian potential has
a small spherically symmetrical  perturbation, analytical expression for the first post-Newtonian approximation
could be obtained following a procedure described in the Landau \& Lifshitz (L \& L) textbook
\cite{Landau_76}.\footnote{In the papers \cite{Dokuchaev_15,Dokuchaev_15a} the authors
evaluated relativistic precessions for a supermassive
black hole case and an additional potential due to a presence of dark matter.}
In paper \cite{Adkins_07}, the authors derived the expression which
is equivalent to the (L \& L) relation in an alternative way and
showed that the expressions are equivalent  and after that they
calculated apocenter shifts for several examples of perturbing
functions.
Assume that a particle moves in slightly perturbed Newtonian potential
\begin{equation}
\Phi_{\rm total}(r)=-\dfrac{GM}{r}+V(r),
\label{potential_total}
\end{equation}
In this case, orbital precession $\Delta\varphi$
per orbital period, induced by small perturbations to the Newtonian
gravitational potential $V(r)<<|\Phi_N(r)|=\dfrac{GM}{r}$ could be expressed as:
\begin{equation}
\Delta\varphi^{rad} = \dfrac{-2L}{GM e^2}\int\limits_{-1}^1
{\dfrac{z
\cdot
dz}{\sqrt{1 - z^2}}\dfrac{dV\left( z \right)}{dz}},
\label{2.1}
\end{equation}
where $L$ is the semilatus
rectum of the orbital ellipse with semi-major axis $a$ and
eccentricity $e$:
\begin{equation}
L = a\left( {1 - {e^2}} \right).
\label{semilatus}
\end{equation}

\subsection{Evaluations of stellar cluster parameters with observations of bright stars near the Galactic Center}

Except a gravitational potential from the supermassive black hole there is an additional gravitational potential formed by an extended mass distribution which is created by a stellar cluster and/or dark matter cloud.  Similarly to  \cite{Gillessen_17} we use
a Plummer profile for an extended mass distribution of a stellar cluster
\begin{equation}
 \rho(r) =  \rho_0 \left(\frac{r}{r_s} \right)^{-\beta},
\label{Plummer}
\end{equation}
where the authors considered $\beta=5/2$ and $\beta=7/4$.
In papers \cite{Dokuchaev_15,Dokuchaev_15a}, the authors used the first post-Newtonian correction
for orbital precession in the case of an extended mass distribution of dark matter. We use the same
ideas to evaluate orbital precession due to presence of a stellar cluster.
Assuming that an extended mass distribution is spherically symmetric, one could consider a distribution
in limits between an apocenter and a pericenter of a selected orbit because a potential of an extended
mass distribution inside a pericenter is equivalent to point mass, while a mass distribution outside apocenter
does note affect a star trajectory. Therefore, similarly to \cite{Dokuchaev_15,Dokuchaev_15a} one has
\begin{equation}
 M_{\rm ext}(r) =  \alpha \left(r^{3-\beta}-r_{\rm peri}^{3-\beta} \right),  \quad r_{\rm peri}< r<  r_{\rm apo},
\quad
 \alpha= \frac{4\pi \rho_0 r_s^{\beta}}{3-\beta},
 \label{Plummer_alpha}
\end{equation}
  $r_{\rm apo}, r_{\rm peri}$ are apocenter and pericenter of a selected orbit, respectively.
Following \cite{Dokuchaev_15,Dokuchaev_15a}, one obtains a perturbing potential consisting of two terms
 \begin{equation}
V_1 (r) = A r^{2-\beta},
\quad
V_2 (r) = \frac{C_1}{r},
 \label{Plummer_potential2}
\end{equation}
where $A=\alpha G$ and $C_1=\alpha G {r_{\rm peri}}^{3-\beta}$.
Substituting the expressions for perturbing potentials into Eq. ({\ref{2.1}) one obtains two integrals
\begin{equation}
\Delta\varphi^{rad}_1 = \dfrac{2AL (\beta-2)}{GM e^2}\int\limits_{-1}^1
{\dfrac{z^{2-\beta}
\cdot
dz}{\sqrt{1 - z^2}}
},
\quad
\Delta\varphi^{rad}_2 = \dfrac{2L C_1}{GM e^2}\int\limits_{-1}^1
{\dfrac{dz}{z\sqrt{1 - z^2}}
},
\label{Adkins_2}
\end{equation}
and a total orbital precession due to a presence of an extended mass distribution formed by a stellar cluster is $\Delta\varphi^{rad}_{\rm tot}=\Delta\varphi^{rad}_1+\Delta\varphi^{rad}_2$.
As it was noted in \cite{Adkins_07,Dokuchaev_15,Dokuchaev_15a,Zakharov_JCAP_18,Zakharov_18}, these integrals  could be expressed through
       the Gauss hypergeometrical function but we leave them as integrals since they look more clear in these forms.

\subsection{Evaluations of black hole parameters with observations of bright stars near the Galactic Center}

Earlier we simulated trajectories of stars in potential formed by black hole and additionally a stellar cluster \cite{Nucita_07},
while in \cite{Zakharov_07} we considered a dark matter component in such a model. We concluded that in the case if a small fraction (around a few percent) of black hole mass is in an extended mass, then trajectories could be significantly different from observed ones.
As it was noted  currently observational data are consistent with a point mass potential and
an extended mass component inside the S2 orbit cannot be more than 1\% of the black hole mass \cite{Gillessen_17}.

\subsection{Constraints on alternative theories of gravity with observations of bright stars near the Galactic Center}

\subsubsection{Graviton mass constraints}


A development of general relativity for more than 100 years was
extremely successful and predictions of GR have been
confirmed with many different experiments and observations.
However, there are many alternative theories and massive theory of gravity
is among the most popular ones.
 A theory of massive gravity was
introduced by M. Fierz and W. Pauli  \cite{Fierz_39}. Later a couple of pathologies of such a gravity theory
have been found, such as the van Dam -- Veltman -- Zakharov --
Iwasaki discontinuity \cite{van_Dam_70,Zakharov_70,Iwasaki_70} for
$m_g \rightarrow 0$ (where $m_g$ is a graviton mass).
Soon after that other pathologies of massive theories of gravity have been found since
Boulware and Deser discovered a presence of ghosts and instabilities in
theory of massive gravity \cite{Boulware_72,Boulware_72b}. However, in the last years
a number of different techniques have been proposed to construct
theories of massive gravity without Boulware -- Deser ghosts
\cite{Rubakov_08}.
A class of ghost-free massive gravity has been proposed in papers
\cite{deRham_10,deRham_11} and such a theory are called now  de Rham --
Gabadadze -- Tolley (dRGT) gravity model (see also reviews \cite{deRham_14,deRham_17}).
A number of different ways to constrain a graviton mass from astronomical
observations are discussed in \cite{deRham_17,Goldhaber_10}. One should note that very often when
people discussed observational constraints on graviton mass they
presented their expectations or forecasts from future observations
since uncertainties and systematics were not carefully analyzed.
Therefore, such estimates are model dependent.

Twenty years ago  C. Will considered an opportunity to
evaluate a graviton mass from observations of gravitational waves
\cite{Will_98} (see also \cite{Will_14} for a more detailed
discussion). Assuming Yukawa gravitational potential of a
form $\propto r^{-1}\exp(-r/\lambda_g)$ \cite{Will_98}
this result indicates that it can be used to constrain the lower
bound for Compton wavelength $\lambda_g$ of the graviton, i.e. the
upper bound for its mass
\begin{equation}
m_{g(upper)}=h\,c/\lambda_g.
\label{mass_vs_lambda}
\end{equation}
The LIGO-Virgo collaboration reported about the first detection of
gravitational waves from a merger of two black holes (it was detected
on September 14, 2015 and it is called GW150914) \cite{Abbott_16}.
Moreover, the team constrained the graviton Compton wavelength
$\lambda_g > 10^{13}$~km which could be interpreted as a constraint
for a graviton mass $m_g < 1.2 \times 10^{-22}$~eV \cite{Abbott_16}
and the estimate roughly  coincides with theoretical predictions
\cite{Will_98}.
 Later the LIGO-Virgo reported the detection of a gravitational wave signal from a
merger of binary black hole system with masses of components
$31.2M_\odot$ and $19.4 M_\odot$ at distance around 880~Mpc which
corresponds to $z \approx 0.18$ \cite{Abbott_17a} and the authors
improved their previous graviton mass constraint since the obtained a new bound $m_g  < 7.7 \times
10^{-23}$~eV \cite{Abbott_17a}.
On August 17, 2017 the LIGO-Virgo collaboration detected a
merger of binary neutron stars with masses around $0.86 M_\odot$ and
$2.26 M_\odot$ at a distance around 40 Mpc (GW170817) and after 1.7 s
the Fermi-GBM detected $\gamma$-ray burst GRB 170817A associated with
the GW170817 \cite{Abbott_17b} meanwhile many other teams detected signatures
of kilonova explosion \cite{Abbott_17c}. The global network Master of robot telescopes played a very important role
in the discovery (see also \cite{Lipunov_17}).
Since gravitational wave
signal was observed before GRB 170817A one could conclude that the
observational data are consistent with massless or very light
graviton, otherwise, electromagnetic signal could be detected before
gravitational one because in the case of relatively heavy gravitons
gravitational waves could propagate slower than light.
Constraints on
speed of gravitational waves have been found
$-3\times 10^{-15} < (v_{g}-c)/c < 7 \times 10^{-16}$   \cite{Abbott_17c}. Graviton
energy is $E=hf$, therefore, assuming a typical LIGO frequency range
$f \in (10,100)$, from the dispersion relation one could obtain a
graviton mass estimate $m_g < 3 \times (10^{-21}-10^{-20})$~eV which
a slightly weaker estimate than previous ones obtained from binary
black hole signals detected by the LIGO team
\cite{Zakharov_IHEP_2017}.

We obtained constraints on Yukawa gravity from observational data on S2 star orbit  \cite{Borka_13}.
Later, we found constraints on graviton mass from these data \cite{Zakharov_16}   (see also discussions
in  \cite{Zakharov_Quarks_16,Zakharov_Baldin_17,Zakharov_MIFI_17,Zakharov_Moriond_17}. In these considerations we used available
data constrain graviton mass. Later, Keck group followed our ideas to improve our estimates with new observational data and the authors obtained $m_g < 1.6 \times 10^{-21}$~eV \cite{Hees_PRL_17}.
In paper \cite{Zakharov_JCAP_18} we considered perspectives to improve a graviton mass estimate with future observational data for S2 and other
bright stars observed with VLT and Keck telescopes, in particular, we  evaluated orbital precession for Yukawa potential and  obtained
an upper limit for a graviton mass assuming that GR prediction about orbital precession will be confirmed with future observations.
As it was shown in \cite{Zakharov_JCAP_18} the longest Compton wavelength could be expressed as
\begin{equation}
\Lambda\approx\dfrac{c}{2}\sqrt{\dfrac{(a\sqrt{1-e^2})^3}{3 G M}}\approx \sqrt{\dfrac{(a \sqrt{1-e^2})^3}{6R_S}},
\label{lambda_delta1}
\end{equation}
or after observations of bright stars for several decades an upper bound for a graviton mass could reach
around $5 \times 10^{-23}$~eV.

\subsubsection{Tidal charge constraints}

An opportunity to evaluate parameters of Reissner -- Nordstr\"om -- de-Sitter  metric from an analysis
of trajectories of bright stars near the Galactic Center is discussed in  \cite{Zakharov_18}.
We use a system of units where $G = c= 1$.
The line element of the spherically symmetric Reissner -- Nordstr\"om -- de-Sitter  metric is
\beq
d s^2 = - f(r) d t^2 + f(r)^{-1} d r^2 + r^2 d\theta^2 + r^2 \sin^2\theta d\phi^2 , \label{metric_RN}
\eeq
where  function $f(r)$ is defined as
\beq
f(r) = 1 - \frac{2M}{r}  + \frac{Q^2}{r^2} - \frac{1}{3} \Lambda r^2. \label{function_f}
\eeq
Here $M$ is a black hole mass, $Q$ is its charge and $\Lambda$ is cosmological constant.
In the case of a tidal charge \cite{Dadhich_01}, $Q^2$ could be negative.
A total shift of a pericenter is \cite{Zakharov_18}
\begin{equation}
  \Delta \theta (total) =   \frac{6\pi M}{L}                 -\frac{\pi Q^2}{ML}         +   \frac{\pi \Lambda a^3 \sqrt{1-e^2}}{M}.
\label{Delta_dS_2}
\end{equation}
and one has a relativistic advance for a tidal charge with $Q^2 < 0$ and apocenter shift dependences on eccentricity and semi-major axis
are the same for GR and Reissner -- Nordstr\"om advance but corresponding factors (${6\pi M}$  and    $-\dfrac{\pi Q^2}{M}$) are different, therefore,
it is very hard to distinguish a presence of a tidal charge and black hole mass evaluation uncertainties.
For $Q^2 > 0$, there is an apocenter shift in the opposite direction in respect to GR advance.


In paper \cite{Zakharov_18} bounds in $Q^2$ and $\Lambda$ are presented for current and future accuracies for Keck and Thirty Meter telescopes.
Following \cite{Zakharov_18} if we adopt   uncertainty  $\sigma_{\rm GRAVITY} = 0.030$~mas for the GRAVITY facilities as it was used in  \cite{Gravity_18} ($\delta_{\rm GRAVITY}=2\sigma_{\rm GRAVITY}$)
or in this case $\Delta \theta (GR)_{S2}= 13.84 \delta_{\rm GRAVITY} $ for S2 star and assuming again that GR predictions
about orbital precession of S2 star will be confirmed  with  $\delta_{\rm GRAVITY}$ accuracy (or $ \left|\dfrac{\pi Q^2}{ML}\right|  \lesssim \delta_{\rm GRAVITY}$) , one could conclude that
\begin{equation}
    |Q^2| \lesssim 0.432 M^2,
\label{RN_GRAVITY}
\end{equation}
or based on results of  future GRAVITY observations one could expect to reduce significantly a possible range of $Q^2$ parameter in comparison with
a possible range of $Q^2$ parameter constrained with current and future Keck data.

\section{Conclusions}
As it was shown monitoring the bright starts at the Galactic Center is very efficient tool to evaluate parameters of gravitational potential
in the framework of GR and also to constrain parameters of alternative theories of gravity as it was shown earlier for massive theory of gravity and for Randall -- Sundrum theory with an extra dimension where Reissner -- Nordstr\"om solution with tidal charge could exist. We also showed an opportunity to constrain $f(R)=R^n$ with such a technique \cite{Borka_12,Zakharov_14}. In the paper we do not discuss an opportunity to observe bright structures around  shadows near supermassive black hole at centers of our Galaxy and M87 with the Event Horizon Telescope\footnote{https://eventhorizontelescope.org/.}to test GR predictions since there are recent reviews on the subject \cite{Zakharov_MIFI_17,Zakharov_18b}.

\subsection*{Acknowledgments}
The author thanks D. Borka, V. Borka Jovanovi\'c,  V. I. Dokuchaev, P. Jovanovi\'c and Z. Stuchl\'ik for useful discussions, NSF
(HRD-0833184) and NASA (NNX09AV07A) at NASA CADRE and NSF CREST
Centers (NCCU, Durham, NC, USA) for a partial support.
The author thanks also the organizers of the International Seminar "Quarks-2018" for their attention to this contribution.


\begin{thebibliography}{200}

\bibitem[\protect\citeauthoryear{Abbott et al.}{2016}]{Abbott_16}
B.~P.~Abbott {\it et al.}, (LIGO Scientific Collaboration and Virgo
Collaboration),
Phys. Rev. Lett. {\bf 116}, 061102 (2016).

\bibitem[\protect\citeauthoryear{Will}{1998}]{Will_98}
C.~Will, 
Phys.  Rev. D  {\bf 57}  2061 (1998); [gr-qc/9709011].







\bibitem{Gillessen_17}
S. Gillessen, P. M. Plewa, F. Eisenhauer et al.,
{Astrophys. J.} \textbf{837}  30, (2017).


\bibitem{Gravity_18}
GRAVITY Collaboration: R. Abuter, A. Amorim, N. Anugu et al.,
{Astron. \& Astrophys. Lett.} \textbf{615}, L15 (2018); arXiv:1807.09409v1[astro-ph.GA].

\bibitem{Genzel_18}
R. Genzel, F. Eisenhauer, S. Gillessen et al.,
ESO1825--Science Release on July 26, 2018,
https://www.eso.org/public/news/eso1825/.

\bibitem[\protect\citeauthoryear{Hees et al.}{2017}]{Hees_PRL_17}
A. Hees, T. Do, A. M. Ghez {\it et al.,}
Phys. Rev. Lett. {\bf 118}, 211101 (2017); arXiv:1705.07902v1 [astro-ph.GA]



\bibitem{Dyson_20}
F. W.	Dyson, A. S. Eddington and C. Davidson,
Phil. Trans.  R. Soc.  London. Series A,
{\bf 220} 291,  (1920).


\bibitem{Kopeikin_99}
S. M. Kopeikin and L. M. Ozernoy, Astrophys. J. {\bf 523}, 771  (1999) .



\bibitem{Alexander_05}
T. Alexander, Phys. Rep. {\bf 419},  65 (2005).

\bibitem{Zucker_06}
S. Zucker, T. Alexander, S. Gillessen et al. Astrophys. J. {\bf 639},  L21 (2006).


\bibitem{Landau_76}
L. D. Landau and E. M. Lifshitz,
\textit{Mechanics}, (Butterworth-Heinemann, Oxford, 1976).

\bibitem{Dokuchaev_15}
V. I. Dokuchaev and Yu. N. Eroshenko,
{ JETP Letters}  {\bf 101},
777  (2015).

\bibitem{Dokuchaev_15a}
V. I. Dokuchaev and Yu. N. Eroshenko, 
{Physics - Uspekhi}  {\bf 58}, 772 (2015).


\bibitem{Adkins_07} G.~S. Adkins and J.~ McDonnell,
{Phys. Rev.
D} \textbf{75}, 082001 (2007) .


\bibitem{Zakharov_JCAP_18}
A. F. Zakharov, P.~Jovanovi{\'c}, D.~Borka,  V.~Borka Jovanovi{\'c},
{J. Cosm.  Astropart.} (JCAP) {\bf 04} (2018) 050; arXiv:1801.04679.

\bibitem{Zakharov_18}
A. F. Zakharov, 
 arXiv:1804.10374[gr-qc].

 \bibitem{Nucita_07}
 A. A. Nucita, F. De Paolis, G. Ingrosso et al.,
{Publ. Astron. Soc.  Pacific} {\bf 119}, 349 (2007).


\bibitem{Zakharov_07}
A.~F.~Zakharov, A.~A.~Nucita, F.~De Paolis and G.~Ingrosso,
Phys. Rev. D {\bf 76},  062001 (2007).

\bibitem[\protect\citeauthoryear{Fierz and Pauli}{1939}]{Fierz_39} M.~Fierz and
W.~Pauli,
{\bf A173},  211  (1939).








\bibitem[\protect\citeauthoryear{Zakharov}{1970}]{Zakharov_70}
V.~I.~Zakharov, 
JETP Letters {\bf 12},  447 (1970).

\bibitem[\protect\citeauthoryear{van Dam \& Veltman}{1970}]{van_Dam_70}
 H.~van Dam and M.~Veltman,  
 Nucl. Phys. B {\bf 22}, 397  (1970) .

\bibitem[\protect\citeauthoryear{Iwasaki}{1970}]{Iwasaki_70}
Y.~Iwasaki, 
Phys. Rev.
D {\bf 2},    2255  (1970).


\bibitem{Boulware_72}
D. G. Boulware and S. Deser,
{Phys. Rev.} {\bf 6},  3368 (1972).

\bibitem{Boulware_72b}
D. G. Boulware and S. Deser,
{Phys. Lett.} B
{\bf 40},  227 (1972).

\bibitem[\protect\citeauthoryear{Rubakov and Tinyakov}{2008}]{Rubakov_08}
V.~A.~Rubakov and P.~G.~Tinyakov, 
Physics -- Uspekhi {\bf 51}, 759 (2008).


\bibitem{deRham_10}
C. de Rham and G. Gabadadze, 
{Phys.Rev.} D {\bf 82}, 044020 (2010).

\bibitem{deRham_11}
C. de Rham,  G. Gabadadze and A. J. Tolley,
{Phys. Rev. Lett.} {\bf 106}, 231101 (2011).

\bibitem{deRham_14}
C. de Rham, 
{Living Rev. Rel.} {\bf 17}, 7
(2014).

\bibitem{deRham_17}
C. de Rham, J. T. Deskins, A. J. Tolley {et al.},
{Rev. Mod. Phys.} {\bf  89}, 025004 (2017).

\bibitem[\protect\citeauthoryear{Goldhaber and Nieto}{2010}]{Goldhaber_10}
A.~S.~Goldhaber and M.~M.~Nieto, 
Rev. Mod. Phys. {\bf 82},   939 (2010).

\bibitem[\protect\citeauthoryear{Will}{2014}]{Will_14}
C.~Will, 
Living Reviews in Relativity  {\bf 17},  4 (2014).



\bibitem{Abbott_17a} B. P. Abbott {et al.},
{Phys. Rev. Lett.} {\bf 118},
221101 (2017).

\bibitem{Abbott_17b} B. P. Abbott {et al.},
{Phys. Rev. Lett.} {\bf 119}, 161101 (2017) .

\bibitem{Abbott_17c} B. P. Abbott {et al.},
{Astrophys. J. Lett.}
{\bf 848},  L13 (2017).

\bibitem{Lipunov_17}
V. M. Lipunov et al., 
{Astrophys. J. Lett.}
{\bf 850},  L1  (2017).

\bibitem{Zakharov_IHEP_2017}
A. Zakharov, P. Jovanovi\'c, D. Borka and V. Borka Jovanovi\'c,
{Intern. J.  Mod. Phys.: Conf. Ser.},
{\bf 47} (2018) 1860096;
arXiv:171208339[gr-qc].


\bibitem{Borka_13} D.~Borka et al.,
J. Cosm. Astropart. Phys. (JCAP) {\bf 11} (2013) 050 .


\bibitem{Zakharov_16}
A.~F.~Zakharov et al.,
J. Cosm. Astropart. Phys. (JCAP) {\bf 05} (2016)  045.

\bibitem{Zakharov_Quarks_16}
A. Zakharov et al.,
{EPJ Web of Conf.} {\bf 125}, 01011 (2016).


\bibitem{Zakharov_MIFI_17}
A. F. Zakharov {et al.},
{Journal of Phys.: Conf. Ser.} {\bf 798}, 012081 (2017).


\bibitem{Zakharov_Baldin_17}
A. F. Zakharov  {et al.},
{EPJ Web of Conf.} {\bf 138}, 010010 (2017).


\bibitem{Zakharov_Moriond_17}
A. F. Zakharov, P.~Jovanovi{\'c}, D.~Borka,  V.~Borka Jovanovi{\'c},
in Proc. the 52nd Rencontres de Moriond -- Gravitation, Eds. E.  Auge, J. Dumarchez, J. Tran Thahn Van,
(ARISF, 2017) p. 247.

\bibitem{Dadhich_01}
N. Dadhich, R. Maartens, Ph.  Papadopoulos and  V.   Rezania, {Phys.
Lett. B} {\bf 487}, 1  (2001).



%
%

\bibitem{Borka_12}
D.~Borka,  P.~Jovanovi\'c,  V.~Borka Jovanovi\'c  et al., 
{Phys. Rev. D}  {\bf 85},   124004 (2012).

\bibitem{Zakharov_14}
A. F. Zakharov, D. Borka, V. Borka Jovanovi\'c, P. Jovanovi\'c,
{Adv.  Space Res.} {\bf 54} (2014) 1108.


\bibitem{Zakharov_MIFI_17}
A. F. Zakharov,
{J.  Phys.: Conf. Ser.} {\bf 934},   012037 (2017).


\bibitem{Zakharov_18b}
A. F. Zakharov, 
{Intern. J. Mod. Phys.} D
{\bf 27},  1841009 (2018).


\end{thebibliography}
\end{document}